\documentclass[twocolumn]{aastex631nl}
\usepackage[fleqn]{amsmath}
\usepackage{float}
\usepackage{lmodern}
\renewcommand{\ng}{NGC\,1365 }
\newcommand{\ngs}{NGC\,1365}
\newcommand{\hst}{{\it HST} }
\newcommand{\hsts}{{\it HST}}
\newcommand{\jwst}{{\it JWST} }
\newcommand{\jwsts}{{\it JWST}}

\newcommand{\hstv}{{\it F555W} }
\newcommand{\hsti}{{\it F814W} }
\newcommand{\hsth}{{\it F160W} }

\newcommand{\jwf}{{\it F200W} }

\newcommand{\hstvs}{{\it F555W}}
\newcommand{\hstis}{{\it F814W}}
\newcommand{\hsths}{{\it F160W}}

\newcommand{\jwfs}{{\it F200W}}

\makeatletter
\newcommand*{\rom}[1]{\expandafter\@slowromancap\romannumeral #1@}
\makeatother
\shorttitle{\ng Cepheids with \jwsts}
\shortauthors{Yuan et al.}

\begin{document}
\title{A First Look at Cepheids in a SN Ia Host with \textbf{\textit{JWST}}}

\author[0000-0001-9420-6525]{Wenlong Yuan}
\affiliation{Department of Physics \& Astronomy, Johns Hopkins University, Baltimore, MD 21218, USA}

\author[0000-0002-6124-1196]{Adam G.~Riess}
\affiliation{Department of Physics \& Astronomy, Johns Hopkins University, Baltimore, MD 21218, USA}
\affiliation{Space Telescope Science Institute, 3700 San Martin Drive, Baltimore, MD 21218, USA}

\author{Stefano Casertano}
\affiliation{Space Telescope Science Institute, 3700 San Martin Drive, Baltimore, MD 21218, USA}

\author[0000-0002-1775-4859]{Lucas M.~Macri}
\affiliation{George P.\ and Cynthia W.\ Mitchell Institute for Fundamental Physics and Astronomy,\\ Department of Physics and Astronomy, Texas A\&M University, College Station, TX 77843, USA}

\begin{abstract}
We report the first look at extragalactic Cepheid variables with the {\it James Webb Space Telescope}, obtained from a serendipitous (to this purpose) observation of NGC 1365, host of an SN Ia (SN 2012fr), a calibration path used to measure the Hubble constant.  As expected, the high-resolution observations with NIRCam through \jwf show better source separation from line-of-sight companions than \hst images at similar near-infrared wavelengths, the spectral region that has been used to mitigate the impact of host dust on distance measurements. Using the standard star P330E as a zeropoint and PSF reference, we photometered 31 previously-known Cepheids in the \jwst field, spanning $1.15 < \log P < 1.75$ including 24 Cepheids in the longer period interval of $1.35 < \log P < 1.75$. We compared the resultant Period-Luminosity relations to that of 49 Cepheids in the full period range including 38 in the longer period range observed with WFC3/IR on \hst and transformed to the \jwst photometric system (\jwfs, Vega). The P-L relations measured with the two space telescopes are in good agreement, with intercepts (at $\log P=1$) of 25.74 $\pm$0.04 and 25.72 $\pm$0.05 for \hst and \jwsts, respectively. Our baseline result comes from the longer period range where the Cepheids have higher signal-to-noise ratios where we find 25.75$\pm 0.05$ and 25.75$\pm 0.06$ mag for \hst and \jwsts, respectively. We find good consistency between this first \jwst measurement and \hsts, and no evidence that \hst Cepheid photometry is ``biased bright'' at the $\sim0.2$ mag level that would be needed to mitigate the Hubble Tension, though comparisons from more SN hosts are warranted and anticipated. We expect future \jwst observations to surpass these in quality as they will be optimized for measuring Cepheids.

\ \par
\end{abstract}

\section{Introduction}

Cepheid variables have held a central role in measuring extragalactic distances for more than a century \citep{1912HarCi.173....1L}.  They exhibit several features which make them uniquely suited for this role.  Their nature is well understood as a consequence of the $\kappa$  mechanism, which drives a periodic overshooting of hydrostatic equilibrium and produces their pulsations \citep{1927MNRAS..87..539E}.  Their great luminosities, $\sim 10^5 L_\odot$, make them visible with modern telescopes at many tens of Megaparsecs.  The large amplitude of their variations uniquely identifies them and their periods standardize their luminosities to a precision of a few percent.  They are ubiquitous in areas of recent star formation, including many hosts of Type Ia supernovae (which have still greater range).  Lastly, hundreds of Cepheids in the Milky Way are in range of precise parallaxes from the ESA {\it Gaia} satellite to provide a 1\% geometric calibration of their fiducial luminosity \citep{2022arXiv220801045R,2022arXiv220809403C}.  For these reasons, Cepheids are the primary distance indicator most often selected for measuring long-range distances and the Hubble constant \citep[][hereafter, R22]{2022ApJ...934L...7R}.

A succession of technological advancements has extended the reach, precision and accuracy of Cepheid distance estimates at tens of Megaparsecs.  One of the original goals of the {\it Hubble} Space Telescope (\hsts) was to resolve extragalactic Cepheids, which was achieved in dozens of galaxies within $\sim$ 20 Mpc with the Wide Field Planetary Camera 2 (WFPC2) at optical wavelengths \citep{2001ApJ...553...47F,2006ApJ...653..843S}.  \hst instruments with greater sensitivity and higher resolution, ACS and WFC3/UVIS, extended this reach to $\sim50$ Mpc and a greater number of nearby SNe~Ia and geometric calibrators \citep{2006ApJ...652.1133M,2011ApJ...730..119R,2016ApJ...830...10H}.  

Given that Cepheids are found in regions of recent star formation, they are observed  through interstellar dust
with a mean reddening (in modestly-inclined spirals, R22) of $E(V-I) \sim 0.3$~mag. Thus, their visible- (0.5$\micron$) and infrared- (0.8$\micron$) band measurements must account for a mean of $\sim0.7$~mag and $\sim0.4$~mag of extinction, respectively, to provide accurate distance measurements, which in consequence are sensitive to the uncertain nature of extragalactic reddening laws.

Wide-scale follow-up of Cepheids in the near-infrared (NIR), to mitigate dust effects, first became practical with WFC3/IR, allowing measurements at $1.6\micron$ and reducing the mean impact of extinction to $\sim$ 0.1 mag and the sensitivity to reddening laws \citep{2011ApJ...730..119R}.  However, the advantage of NIR observations over optical bands came with new challenges; at these wavelengths, the resolution of \hst is 2-3 times lower and the background (in the form of ubiquitous red giants) is an order of magnitude greater.  The result is an increase in the measurement errors (after {\it statistical}  removal of the backgrounds measured using artificial stars) which may limit the precision of distance measurements without a large number ($>$50) of Cepheids in each host.  While Cepheid distance measurements from either the optical or NIR are in good agreement (R22), a result most likely if both are accurate, the pursuit of a 1\% measurement of the Hubble constant demands ever more stringent tests of Cepheid photometry. 

The newly-launched {\it James Webb} Space Telescope (\jwsts) offers the twin advantages of angular resolution comparable to WFC3/UVIS at visible wavelengths and the lower impact of interstellar dust as WFC3/IR in the same observation. \jwst observations planned for its first GO cycle have been designed to take advantage of these capabilities and reobserve Cepheids previously measured with \hsts, work which is likely to require years to collect and thoroughly analyze to fully come to fruition.  However, an early observation with \jwst of a SN Ia host previously observed by \hst offers a serendipitous (for this endeavor) and valuable preview.

To be clear in setting expectations for future \jwst observations, these {\it serendipitous} observations of the Cepheids in \ng fall short of demonstrating the full capability of the observatory for this endeavor. They are shorter in exposure time by a factor of a few than those planned for this purpose and they are obtained at nearly twice the wavelength needed to optimally resolve and reduce the contributions of nearby red giants (i.e., the background).  Notably, they cover a more crowded region along a spiral arm (see Figure~\ref{fig_obs}) compared to most of those observed by \hsts.  Further, they provide only a single (i.e., ``random'') epoch or phase in each Cepheid light curve, which adds an additional dispersion of $0.1$ to $0.2$ mag depending on the amplitude of the Cepheid. Lastly, the state of the \jwst calibration data (e.g., flat fields, dark frames, bias frames, geometric distortion maps, linearity corrections) is in its first iteration and will improve with time. Nevertheless, and with these limitations in mind, these observations preview the enhanced capabilities of \jwst over \hst and provide meaningful, if preliminary, quantitative results.

In \S2 we describe the details of the \jwst observations for \ngs, as well as the data reduction and photometry procedures. We show our results in \S3 and a brief discussion in \S4.  An appendix provides information about past \hst observations of Cepheids in \ng for easy reference.

\section{Observations, data reduction, and photometry}

\subsection{Observations \& data reduction}

\begin{figure*}
\epsscale{1.2}
\plotone{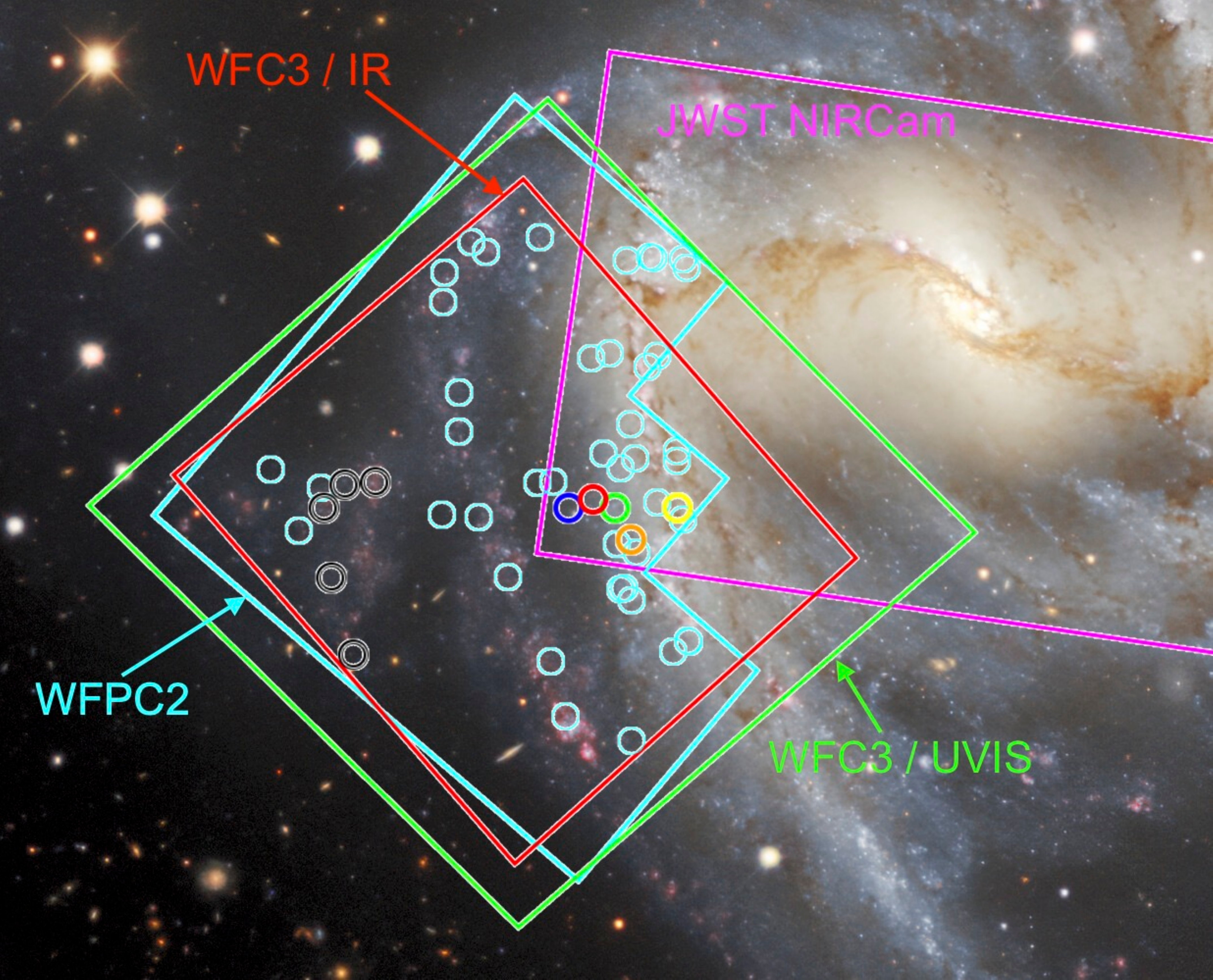}
\caption{Observation footprints of \ng with \jwst NIRCam (magenta), \hst WFPC2 (cyan), WFC3/UVIS (green), and WFC3/IR (red) overlaid on a color composite image from the Dark Energy Survey (DOE/FNAL/DECam/ CTIO/NOIRLab/NSF/AURA). The locations of Cepheids used in this study are indicated by circles. North is up and east is to the left.}\label{fig_obs}
\end{figure*}

The central region of \ng was recently observed with \jwst NIRCam on 2022 August 13 as part of program GO-2107 (PI: Janice Lee), which aims to study the star formation activity in 19 nearby galaxies. The \ng field partially overlaps with an \hst WFPC2 time-series field (GO-5972, PI: Jeremy Mould) where dozens of Cepheids were discovered \citep{1999ApJ...515....1S,2016ApJ...830...10H} and followed up in the NIR (R22). With the Cepheid locations and periods determined from those \hst data, we have an opportunity to photometer and study these Cepheids in the new \jwst observations. In Figure~\ref{fig_obs} we show the footprints of the \jwst observations as well as archival \hst observations and locations of previously-identified Cepheids. The initial WFPC2 time-series and WFC3 follow-up targeted a less crowded part of the host off the spiral arms, but the NIRCam observations targeted the center of the galaxy  and primarily contain Cepheids in a small dense, crowded region. Appendix Figure~\ref{fig_a} shows less-crowded Cepheids imaged by \hst that are more similar to those typically studied in \hst fields.   Due to the overlap of the two observatories, we can also directly compare the images and measurements of many of the  same Cepheids in the denser regions of the host.

\ \par

We retrieved \jwst observations of \ng from MAST and processed the raw data (stage 0) using the {\tt JWST} Science Calibration Pipeline version 1.6.2. There are 25 exposures in total, with the short-wavelength channel through the \jwf filter and the long-wavelength channel through the {\it F300M}, {\it F335M}, and {\it F360M} filters. In this study, we only analyzed the \jwf data for their depth and proximity in wavelength coverage compared to the \hst \hsth band. The \jwf data consist of eight subfields, with each one covered by approximately one short-wavelength detector. Only the two east-most subfields contain previously-identified Cepheids; thus, we excluded the other six from the analysis. The total exposure times are 1202.52s for both analyzed subfields.

We noticed the 1/f noise causing small bias shifts in the calibrated stage 2 data products \citep[see \S2 of][]{2022arXiv220711701M}. We corrected them by subtracting the median value of each row and then each column before the {\tt JWST} pipeline stage 3 process. Similar to \citet{2022arXiv220711701M}, we masked all  sources when computing the median values for row and column subtractions.

\begin{figure*}
\epsscale{1.05}
\plotone{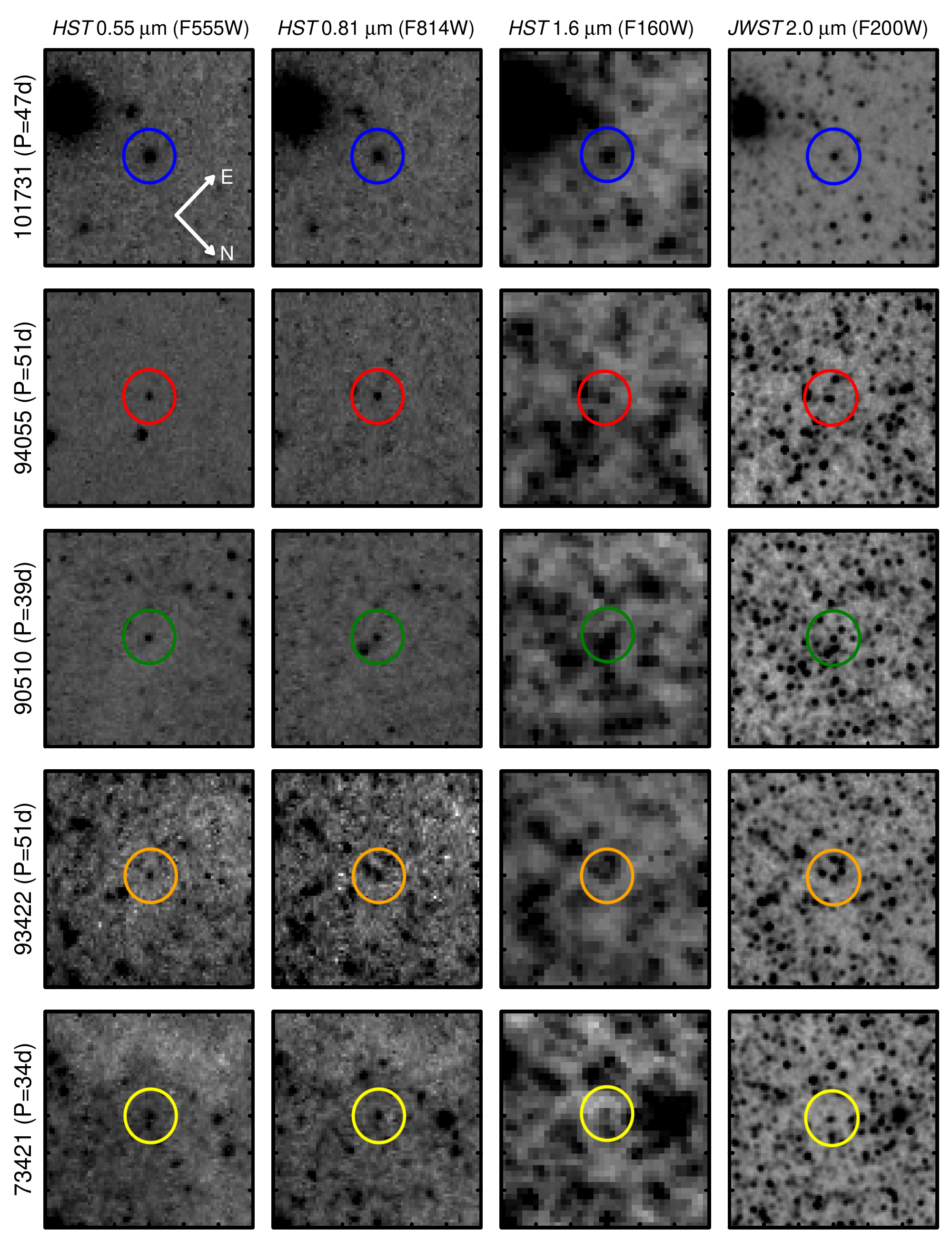}
\caption{Image cuts of 5 example Cepheids analyzed in this study. Their locations are indicated by the corresponding colors in Figure~\ref{fig_obs}. The circles cover a radius of 0\farcs375\ while the image cuts display 3\arcsec\ in a side. From left to right, each row shows one (same) Cepheid in \hst \hstvs, \hstis, \hsths, and \jwst \jwfs, where the exposure times are 1410s, 1770s, 3618s, and 1203s, respectively. The orientation of the image cuts is indicated by the white compass in the top-left panel.}\label{fig_stamp}
\end{figure*}

We used the WCS in the images to locate the Cepheids based on their \hst positions.  We identified a global shift of $\sim0\farcs5$ between the \hst and \jwst positions and accounted for this to register the images.  After this global shift we found point sources at the expected positions of the Cepheids to a precision of less than a NIRCam pixel ($0\farcs031$; see Figure~\ref{fig_stamp}).  The \hst NIR observations in these spiral arms are under-sampled (even after drizzling to $0\farcs08/$pixel resolution) and lack the inherent resolution of \jwst (despite the greater wavelength of those observations). 

While the Cepheids were easily apparent in the deeper and higher-resolution images in \jwfs, they were hard to discern in the accompanying observations at longer-wavelengths and through medium-width bands due to their much shorter exposure times, lower angular resolution and  lower throughput of these filters.  As a result we only analyzed the \jwf images.

\subsection{Photometry}

We performed point-spread function (PSF) photometry using a crowded-field photometry package based on DAOPHOT/ALLSTAR \citet{1987PASP...99..191S,1994PASP..106..250S}. We constructed an empirical model of the PSF using \jwf observations of the standard star P330E  (taken on 2022 Aug 29, obs.~ID=jw01538o155t002) obtained in a 160-pixel subarray (using a minimal exposure time to keep the star below saturation) which included two dithers placed on each of the B-module chips.  

We chose not to use the pipeline calibration to obtain the image zeropoints as they have been found to have limited accuracy (at the time of this writing) including  chip-to-chip offsets \citep[and possible time-dependence between the early life of the mission and the present,][]{Brammer:2022,Boyer:2022,Nardiello:2022}.  To produce reliable zeropoints for the observation of NGC 1365 we used the above observations of P330E obtained and combined for each B-module chip separately to directly calibrate the Cepheids observed in that chip.  We assigned each image of P330E a reference Vega magnitude of 11.42 mag \citep{2022AJ....163...45R}.  An important advantage of using the Aug 29, 2022 observations of P330E to set the zeropoints for the images of NGC 1365 is that they were obtained only 2 weeks after the observation of NGC 1365, an interval during which JWST's wave front monitoring has shown it to be relatively stable with modeled photometric variations over the interval of $<$ 0.01 mag (M. Perrin, 2022 private communication).  (We did not make use of aperture photometry for the Cepheids due to the inability to separate nearby sources as expected from inspection of Figure~\ref{fig_stamp}.)

To avoid a flux bias from the determination of Cepheid positions in \hst NIR images, it is necessary to fix their locations using the uncrowded optical images \citep[i.e., ``forced photometry'', ][]{2009ApJ...699..539R}. The algorithm fits the PSF of the Cepheids at their known, fixed positions, subtracts them from the images, identifies additional, unresolved sources down to a fixed threshold, and then simultaneously optimizes the fit to the non-Cepheids (parameters are x, y and flux) and Cepheids (parameter is flux) to determine the latter's flux.  We then add ``artificial stars'' at the same brightness as the Cepheid (based on the period and iterative fit of the Period-Luminosity relation), and remeasure these using the same procedure to account for the mean background of unresolved sources near the position of the Cepheid (i.e., a statistical crowding correction) and to measure the uncertainty in the Cepheid magnitude. We also compared our results to the level 3, full-calibrated images produced by the STScI pipeline and found that the photometry was consistent between the versions of the images.

\begin{figure*}
\begin{center}
\includegraphics[width=0.81\textwidth]{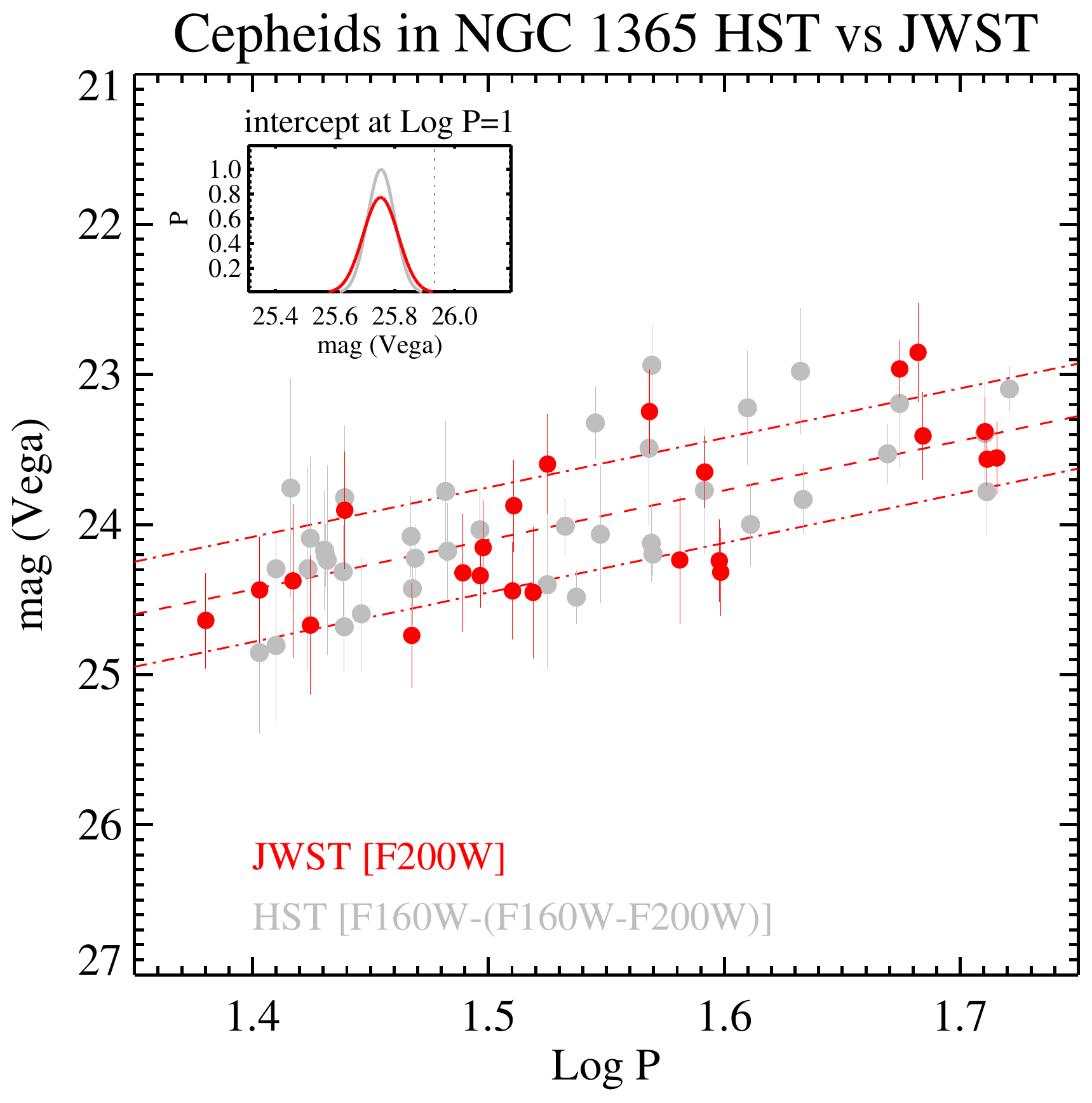}
\end{center}
\caption{Near-infrared Period-Luminosity relations for Cepheids in the range $1.35<\log P <1.75$ (baseline results) measured with \hst and \jwsts.  The \jwst sample (red) includes 24 Cepheids observed in \jwf ($2\micron$).  The \hst sample includes 38 Cepheids from R22 with \hsth magnitudes transformed to \jwf using a color transformation based on their measured $V-I$ colors and \hsths$-$\jwfs.  The inset shows the intercepts of the relations at $\log P = 1$.  The solid red curve uses the \jwst PSF photometry calibrated to P330E.}\label{fig_plr}
\end{figure*}

\section{Results}

Fixing the slope of the Period-Luminosity relation to the global value of $-3.30$ determined from the mean of thousands of Cepheids in the MW, LMC, SMC, M31, NGC\,4258 and SN Ia hosts in the NIR (R22), we measured the intercepts at $\log P=1$. 

For our ``baseline'', we limited the comparison to a period range of $1.35 < \log P < 1.75$ where the Cepheids as measured from both telescopes have strong signal-to-noise ratios.  Below this range the SNR at $F160W=24.5$ (Vega) drops to $<$10 and above this range Cepheid periods in \ng are not expected to be accurate because the original time-series used to find the Cepheids in \ng spanned only 48 days ($\log P=1.68$), so that a full cycle would not have been seen.  The \jwst and \hst Cepheid Period-Luminosity relations are shown in Figure~\ref{fig_plr}.  

For \jwst with PSF fitting (referenced to P330E) and with 24 Cepheids we find an intercept of 25.75$\pm 0.06$ (SD=0.36 mag).  In Table \ref{tab_sam} we provide intercepts for broader ranges of periods and with and without $\sigma$ clipping.

To directly compare the \hst and \jwst Period-Luminosity relations observed at different, though adjacent, bandpasses, it is necessary to account for their different wavelength responses.  Due to the simple spectral energy distributions of stars, particularly on the Rayleigh-Jeans tail in the NIR, it is relatively straightforward to estimate this difference, which is the color \hsths$-$\jwfs, from another measured color such as \hstvs$-$\hstis.  To do this rigorously we used the PARSEC isochrones \citep{2012MNRAS.427..127B} for stellar atmospheres which are provided as calculated for the \hst and \jwst bandpasses (using version CMD v3.6, \url{http://stev.oapd.inaf.it/cgi-bin/cmd}).  We limited these to a range appropriate for Cepheids: ages of 10 to 100 Myr, $T_{\rm eff}$ of 4000 to 7000 degrees, initial masses $>2M_\odot$, and $\log g <$ 2.  These stars have a tight locus in the color-color plane of WFC3/UVIS for 
\hstvs$-$\hsti vs \hsth(\hsts)$-$\jwf(\jwsts). We fit a second-order polynomial to the color-color relation, finding 

\vspace{-0.6cm}
{\setlength{\mathindent}{0cm}
\begin{align*}
\hsth - \jwf = 0.007 & + 0.053 (\hstv - \hstis) \\
                     & + 0.077 (\hstv - \hstis)^2
\end{align*}
}
\vfill

\begin{deluxetable*}{rcccccc}[ht]
\tablecaption{\jwst \jwf Cepheid Photometry \label{tab_cep}}
\tablewidth{0pt}
\tablehead{
\colhead{ID} & \colhead{$P$} & \colhead{\jwf$^a$} & \colhead{$\sigma^b$} & \colhead{R.A.$^c$} & \colhead{Decl.} & {subfield}\\ 
& [days] & \multicolumn{2}{c}{[mag]} & \multicolumn{2}{c}{[deg] (J2000.0)} & 
}
\startdata
    97917 &     24.00 &     24.64 &      0.32 &    53.433156 &   -36.158061 &    south \\
    60205 &     25.30 &     24.44 &      0.36 &    53.435680 &   -36.144208 &    south \\
    25668 &     26.13 &     24.38 &      0.51 &    53.432499 &   -36.136873 &    north \\
    74699 &     26.58 &     24.67 &      0.46 &    53.427052 &   -36.156193 &    south \\
    40364 &     27.48 &     23.90 &      0.39 &    53.429580 &   -36.143996 &    south \\
    65664 &     29.34 &     24.74 &      0.35 &    53.431935 &   -36.149101 &    south \\
    53380 &     30.85 &     24.32 &      0.40 &    53.433946 &   -36.143847 &    south \\
   100027 &     31.37 &     24.34 &      0.21 &    53.439120 &   -36.153423 &    south \\
    79315 &     31.46 &     24.15 &      0.31 &    53.432884 &   -36.152400 &    south \\
    80300 &     32.38 &     24.44 &      0.32 &    53.434414 &   -36.151327 &    south \\
    94995 &     32.42 &     23.87 &      0.31 &    53.431335 &   -36.158716 &    south \\
    45761 &     33.03 &     24.45 &      0.44 &    53.430433 &   -36.144804 &    south \\
    73421 &     33.50 &     23.60 &      0.33 &    53.427504 &   -36.155372 &    south \\
    61628 &     37.01 &     23.25 &      0.28 &    53.427674 &   -36.151793 &    south \\
    17203 &     38.12 &     24.24 &      0.43 &    53.430027 &   -36.136683 &    north \\
    90510 &     39.06 &     23.65 &      0.24 &    53.433311 &   -36.155466 &    south \\
    77265 &     39.61 &     24.24 &      0.27 &    53.429382 &   -36.154908 &    south \\
    58983 &     39.67 &     24.32 &      0.29 &    53.427627 &   -36.151066 &    south \\
   101731 &     47.24 &     22.96 &      0.19 &    53.437542 &   -36.155483 &    south \\
     8616 &     48.09 &     22.85 &      0.33 &    53.427309 &   -36.136727 &    north \\
     9712 &     48.33 &     23.41 &      0.29 &    53.426932 &   -36.137379 &    north \\
    93422 &     51.34 &     23.38 &      0.23 &    53.431778 &   -36.157790 &    south \\
    94055 &     51.45 &     23.56 &      0.18 &    53.435355 &   -36.154809 &    south \\
    17544 &     51.94 &     23.56 &      0.24 &    53.430280 &   -36.136566 &    north
\enddata
\tablecomments{
$a$: These are Vega mag referenced to P330E = 11.42 in \jwfs. $b$: The errors are derived from artificial stars and also include a random phase error in quadrature of 0.15 mag. $c$: Positions are referenced to the WCS of \jwst images processed using {\tt JWST} pipeline v1.6.2.}
\end{deluxetable*}

\begin{deluxetable*}{lccc}[h]
\tablecaption{\hst and \jwst Intercepts at $\log P=1$ (slope=$-3.30$) for NIR Cepheids in \ngs \label{tab_sam}}
\tablewidth{0pt}
\tablehead{
\colhead{Sample} & \colhead{$N$ Cepheids} & \colhead{Period range} & \colhead{\jwf Intercept$^a$}
}
\startdata
{\bf \pmb \hst WFC3/IR field, baseline} & {\bf 38} & {\bf 1.35 $< \pmb \log~\pmb P <$ 1.75} & {\bf 25.754 $\pmb\pm\pmb{0.045}$} \\
\hst WFC3/IR field, extended & 49 & 1.15 $< \log P <$ 1.75 & 25.736 $\pm 0.043$ \\
\hst WFC3/IR field, SH0ES R22$^b$ & 46 & 15.0 $< P <$ 50.0 & 25.750 $\pm 0.045$ \\
\hline
{\bf \pmb \jwst NIRCam field, baseline, PSF} & {\bf 24} & {\bf 1.35 $< \pmb \log~\pmb P <$ 1.75} & {\bf 25.752 $\pmb\pm\pmb{0.059}$} \\
\jwst NIRCam field, extended, PSF & 31 & 1.15 $< \log P <$ 1.75 & 25.718 $\pm 0.055$ \\
\enddata
\tablecomments{ $a$: Results from \hst measured in \hsth and converted to \jwf using
$\hsth - \jwf = 0.007 + 0.053 (\hstv - \hstis) + 0.077 (\hstv - \hstis)^2$. 
$b$: Same period range and sample used in R22.}
\end{deluxetable*}
\clearpage

The dispersion of the synthetic values around this approximation is 0.007 mag.  The mean Cepheid color of the sample is $\hstv-\hsti=1.08$~mag (sample SD=0.22 mag) where the relation gives $\hsth-\jwf$=0.15 mag (sample SD=0.05 mag), however we computed the individual values for each Cepheid, as given in the Appendix. We subtract the individual \hsths$-$\jwf colors predicted from the optical colors from the measured \hst \hsth to provide a direct comparison to \jwst \jwf as shown in Figure~\ref{fig_plr}.

The baseline measurements of the \hst intercepts use the \hsth magnitudes as given in R22, the \hsths$-$\jwf colors as given in the Appendix, and include 38 Cepheids in this period range. To increase the sample for the purpose of this \hst to \jwst comparison, we added 3 Cepheids with $P=51,51,$ and 52 days found by \citet{2016ApJ...830...10H} and only slightly above the $P<50$ day limit used by R22 but still well below the $1.2\times$ time-span of the observations necessary to be reliable. We find an intercept for \hst at $\log P=1$ of 25.75$\pm 0.05$ mag and provide intercepts with other period ranges in Table \ref{tab_sam}.  

The inset in Figure~\ref{fig_plr} compares the intercepts. The agreement between the \hst and \jwst intercepts is very good, below 1$\sigma$ in their difference.  The same mean difference is seen when comparing only identical Cepheids though the number of Cepheids measured by both is far smaller and thus the comparison is less significant.  The dispersion around the Period-Luminosity relation as shown in Figure~\ref{fig_plr} is comparable between \hst and \jwst and is likely to be smaller for optimal \jwst observations with multiple epochs, better image calibration and in less crowded regions more typically observed with \hsts.

 To a $\sim$0.05 mag level of preliminary accuracy based on still limited characterization of \jwst and for this case we can conclude that past \hst NIR measurements do not appear biased, let alone ``biased bright'' at the $\sim$0.2 mag level (i.e., by the systematics of past photometry measurements or by previously unresolved companions) as could mitigate the ``Hubble Tension'' in R22 (and then only if such a bias was not also similarly present in \hst  photometry of Cepheids in the geometric anchor, NGC$\,$4258).  


\section{Discussion}

The \jwst images and measurements of Cepheids in \ng and in comparison to those from \hst bode well for the quality of such future measurements.  We reiterate that these observations were not optimized for observing Cepheids and are far from the best that \jwst can do.  Optimal observations would be longer in exposure time, cover multiple passbands to the necessary depth, include shorter wavelengths for better resolution, include multiple epochs to reduce the random phase noise, have higher signal-to-noise calibration frames (flats, darks, bias frames, chip offsets, geometric distortion for locating Cepheids, etc) available and better cover the regions where past \hst programs have found Cepheids and measured their periods. 

We also note that it is too early in the life of \jwst and NIRCam to identify and calibrate subtle photometric effects.  There is one such effect we are aware of, the count-rate non-linearity (CRNL), which makes faint objects appear fainter, though the scale of this effect has been diminishing with improvements in NIR detector manufacturing and testing used to select the best chips.  Because the level of CRNL has not yet been measured in space for NIRCam, we did not correct either the NIRCam or the WFC3/IR Cepheid photometry for this effect, so to first approximation we might expect that CRNL cancels in the comparisons provided here.  For WFC3/IR, CRNL makes the Cepheids in \ng $\sim 0.03$ mag faint relative to the flux level of standard stars \citep{2009ApJ...699..539R}.  If the CRNL of NIRCam is $\sim$ half the level of WFC3/IR (our guess), the error in the comparison will be $\sim 0.015$ mag, negligible at the precision of {\it this} study, but important to calibrate for future, larger samples. The single-epoch sampling of this \jwst observation introduces a statistical bias of $\sim 0.005$~mag in the Cepheid Period-Luminosity relation compared to the typical flux-averaged (multi-epoch) observations. This bias is again negligible for the precision of this study.

Nevertheless, the quantitative comparison of the first \jwst Cepheid Period-Luminosity intercepts presented here is promising, and already significant as a check on past \hst measurements.   We expect that the calibration of this observatory will only improve and mature, leading to future observations that should provide ever more definitive investigations.

\begin{acknowledgments}
We are indebted to all of those who spent years and even decades bringing \jwst to fruition.  We are grateful to the proposers of GO-2107 (PI: Janice Lee) for making their program non-proprietary, enabling the community to undertake assorted investigations from this data including this study. This research made use of the NASA’s Astrophysics Data System.
\end{acknowledgments}

\clearpage

\bibliographystyle{aasjournal}
\bibliography{ref}

\appendix
\restartappendixnumbering
\section{Cepheid measurements from \hsts}
\ \par

\begin{deluxetable*}{rrrrccrr}[b]
\tablecaption{\hst \hsth Cepheid Photometry \label{tab_hcep}}
\tabletypesize{\footnotesize}
\tablewidth{0pt}
\tablehead{
\colhead{ID} & \colhead{$P$} & \colhead{\hsth} & \colhead{$\sigma$} & \colhead{\hstvs$-$} & \colhead{\hsths$-$} & \colhead{R.A.$^a$} & \colhead{Decl.}\\[-6pt]
\colhead{} & \colhead{} & \colhead{} & \colhead{} & \colhead{\hstis} & \colhead{\jwfs} & \colhead{} & \colhead{}\\[-6pt]
& \colhead{[days]} & \multicolumn{4}{c}{[mag]} & \multicolumn{2}{c}{[deg] (J2000.0)}}
\startdata
    60205 &     25.16 &     25.03 &      0.54 &      1.18 &      0.18 &    53.435572 &   -36.144146 \\
   136735 &     25.57 &     24.41 &      0.23 &      0.89 &      0.12 &    53.465135 &   -36.152743 \\
    43927 &     25.57 &     24.98 &      0.50 &      1.17 &      0.17 &    53.440450 &   -36.135135 \\
   101154 &     25.94 &     23.89 &      0.73 &      0.98 &      0.13 &    53.426225 &   -36.165263 \\
   106082 &     26.38 &     24.40 &      0.68 &      0.84 &      0.11 &    53.432670 &   -36.161386 \\
    74699 &     26.44 &     24.28 &      0.55 &      1.23 &      0.19 &    53.426941 &   -36.156136 \\
    63449 &     26.81 &     24.39 &      0.40 &      1.35 &      0.22 &    53.445400 &   -36.136227 \\
   138773 &     26.83 &     24.35 &      0.21 &      1.04 &      0.15 &    53.462525 &   -36.157297 \\
   101112 &     26.88 &     24.37 &      0.63 &      0.98 &      0.13 &    53.426292 &   -36.165183 \\
   120972 &     27.30 &     24.46 &      0.31 &      1.04 &      0.15 &    53.443078 &   -36.160625 \\
   126914 &     27.33 &     24.78 &      0.30 &      0.80 &      0.10 &    53.455397 &   -36.153646 \\
    40364 &     27.34 &     23.94 &      0.48 &      0.91 &      0.12 &    53.429471 &   -36.143937 \\
    65336 &     27.79 &     24.71 &      0.38 &      0.89 &      0.12 &    53.446800 &   -36.135500 \\
   124631 &     29.17 &     24.22 &      0.27 &      1.02 &      0.14 &    53.438970 &   -36.166817 \\
   130859 &     29.21 &     24.56 &      0.24 &      0.98 &      0.13 &    53.458170 &   -36.153817 \\
   133465 &     29.29 &     24.44 &      0.23 &      1.34 &      0.22 &    53.460423 &   -36.154001 \\
   105797 &     30.17 &     23.98 &      0.47 &      1.28 &      0.20 &    53.431618 &   -36.162206 \\
   106470 &     30.23 &     24.30 &      0.32 &      0.93 &      0.12 &    53.427648 &   -36.166054 \\
   100027 &     31.20 &     24.11 &      0.29 &      0.66 &      0.08 &    53.439011 &   -36.153369 \\
    73421 &     33.32 &     24.52 &      0.56 &      0.91 &      0.12 &    53.427399 &   -36.155314 \\
   122163 &     33.91 &     24.20 &      0.19 &      1.23 &      0.19 &    53.449210 &   -36.155957 \\
   139368 &     34.28 &     24.64 &      0.18 &      1.09 &      0.16 &    53.459369 &   -36.160824 \\
    87703 &     34.92 &     23.51 &      0.24 &      1.22 &      0.19 &    53.449343 &   -36.140061 \\
   117850 &     35.09 &     24.21 &      0.46 &      1.04 &      0.15 &    53.445830 &   -36.156138 \\
    61628 &     36.80 &     23.79 &      0.52 &      1.63 &      0.30 &    53.427562 &   -36.151741 \\
   103387 &     36.88 &     24.25 &      0.26 &      0.94 &      0.12 &    53.447790 &   -36.146777 \\
    80315 &     36.90 &     23.12 &      0.27 &      1.20 &      0.18 &    53.449213 &   -36.137887 \\
   142648 &     36.94 &     24.32 &      0.14 &      0.93 &      0.12 &    53.457318 &   -36.166545 \\
    90510 &     38.84 &     23.92 &      0.42 &      1.05 &      0.15 &    53.433201 &   -36.155411 \\
   128912 &     40.51 &     23.39 &      0.38 &      1.14 &      0.17 &    53.437587 &   -36.170953 \\
   103704 &     40.63 &     24.24 &      0.28 &      1.43 &      0.24 &    53.440403 &   -36.153516 \\
   104907 &     42.66 &     23.14 &      0.42 &      1.11 &      0.16 &    53.432245 &   -36.161310 \\
   123489 &     42.77 &     23.96 &      0.23 &      0.95 &      0.13 &    53.431458 &   -36.172785 \\
   109560 &     46.44 &     23.80 &      0.20 &      1.54 &      0.27 &    53.447707 &   -36.149730 \\
   101731 &     46.99 &     23.31 &      0.43 &      0.90 &      0.12 &    53.437427 &   -36.155425 \\
    93422 &     51.07 &     23.65 &      0.37 &      1.49 &      0.26 &    53.431666 &   -36.157737 \\
    94055 &     51.17 &     24.00 &      0.29 &      1.35 &      0.22 &    53.435250 &   -36.154750 \\
   134975 &     52.31 &     23.21 &      0.15 &      0.88 &      0.11 &    53.460099 &   -36.155567
\enddata
\tablecomments{
$a$: Positions are referenced to the WCS of \hst \hsth images processed using {\tt AstroDrizzle} v2.2.6.
}
\end{deluxetable*}

\begin{figure*}
\epsscale{0.8}
\plotone{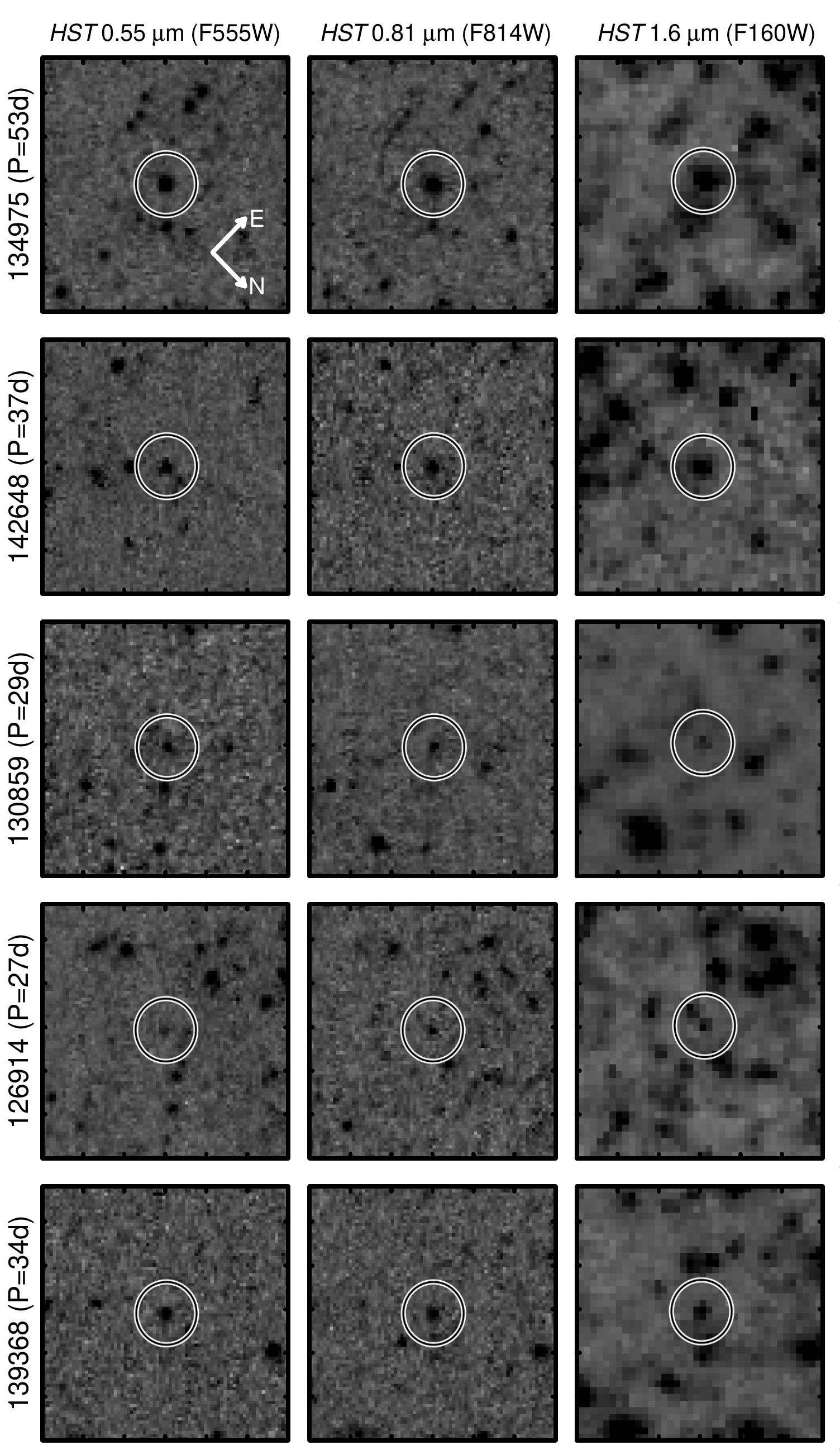}
\caption{Same as Figure~\ref{fig_stamp} but for 5 examples in the more typical, lower-density \hst \hsth field, where \jwst observations are not available.  See Figure~\ref{fig_obs} for location.}\label{fig_a}
\end{figure*}

\end{document}